\documentclass[journal]{IEEEtran}
\usepackage{ifpdf}
\usepackage{cite}
\ifCLASSINFOpdf
\usepackage[pdftex]{graphicx}
\else
\usepackage[dvips]{graphicx}
\fi
\usepackage{amsmath}
\usepackage{algorithmic}
\usepackage{array}
\usepackage{url}
\usepackage{color}
\begin{document}
\title{An Efficient Ionization Method for Pressure Up To Thousands of Pascals}
\author{Lei Chang and Xinyue Hu
\thanks{Authors are with the School of Aeronautics and Astronautics, Sichuan University, Chengdu,
Sichuan Province, 610065, P. R. China, e-mail address: leichang@scu.edu.cn.}}
\maketitle

\begin{abstract}
A novel radio-frequency antenna consisting of isolated solenoids is reported to generate plasma for pressure up to thousands of pascals. The underlying mechanism involves capacitive coupling, inductive coupling and axial confinement from magnetic mirrors. It can ionize the air reliably with pressure up to $2500$~Pa for the $1$~kW-$13.56$~MHz power supply employed, and maintain the discharge steadily from tens of minutes to hours. This ionization method is particularly useful for plasma propulsion in near space and other low-temperature plasma applications where sub-atmospheric pressure is required. 
\end{abstract}

\IEEEpeerreviewmaketitle

It has been an interesting question whether the ion engines or plasma thrusters, which are widely used in outer space, can be employed on the aircrafts for atmospheric environment. This is usually challenging for traditional aeronautic space, because the efficiency of momentum transference between ions and neutral air molecules is very low and the power consumption for ionization is extremely high for atmospheric pressure at low altitudes[1-5]. However, for the emerging field of near space with altitude of $20-100$~kM and corresponding pressure in range of $5529-0.032$~Pa[6], it becomes possible to explore novel plasma propulsion technique as main thruster for supersonic aircrafts, and the primary focus is efficiently producing plasma for thousands of pascals. 

There are few ionization methods which can work from nearly vacuum to atmospheric pressure, and dielectric barrier discharge is one of them[7-9]. However, the short distance between electrodes and low breakdown voltage of dielectric material limit the production of large-volume plasma, long-life steady discharge and total momentum exerted on the ambient air. Inspired by the easy ignition of capacitively coupled plasma under high pressure and remarkable ionization efficiency of inductively coupled discharge without contact between electrode and plasma[10], we propose a novel radio-frequency antenna based on hybrid capacitive and inductive coupling effects and confinement from magnetic mirrors. 

A schematic of the designed antenna is shown in Fig.~$1$, together with the layout of experimental setup. The antenna is a broken solenoid, which results in two parts separated by a controllable distance. This separation is intended to introduce a large voltage difference in the axial direction for ion acceleration, but also contributes to the ionization process. Different from the separated metal sleeves, which generate purely capacitive discharge, these isolated solenoids produce plasma inductively underneath each coil and capacitively between the axial coils. Therefore, the ionization procedure composes of both inductive and capacitive coupling effects, and they are coupled to each other during the whole discharge. Moreover, the strong magnetic field induced by each coil behaves as a magnetic mirror, so that the formed plasma is somewhat confined between these two mirrors, which also enhances the ionization efficiency and is in favor of long-time steady discharge. 
\begin{figure}[ht]
\begin{center}
\includegraphics[width=0.49\textwidth,angle=0]{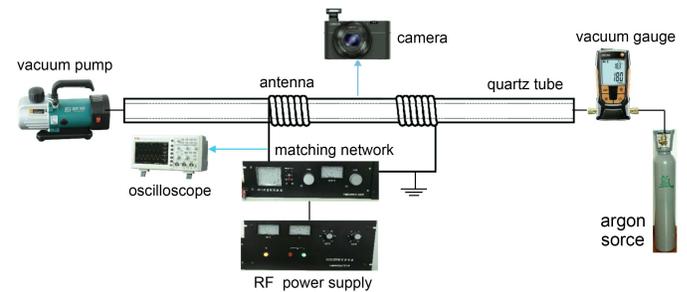}
\end{center}
\caption{Schematic of the hybrid capacitive and inductive coupling antenna and layout of experimental setup.}
\label{fg1}
\end{figure}
The antenna is connected to a radio-frequency power supply ($1$~kW-$13.56$~MHz) through a matching network, and wrapped outside a cylindrical quartz tube. The tube connects a vacuum pump to a bi-pass vacuum gauge, which measures the pressure between the discharge chamber (quartz tube) and argon source. It should be noted that the argon was used only as an initial test of the antenna, and all results presented here are based on air only. We switch off the argon source completely after proving that the designed antenna can work for a wide range of pressure with radio-frequency power supply, and refill the tube with common air before conducting the presented experiment here. We record the discharge image by a digital camera and voltage signal by an oscilloscope. 

The experiment was carried out with pressure decreasing gradually from $1$~atm to tens of pascals and then increasing back to $1$~atm. The vacuum gauge could measure the real-time pressure accurately in range of $0-2500$~Pa, while the vacuum pump could only achieve tens of pascals within an hour. We found that the discharge started to occur when the pressure dropped below $1500$~Pa, but could sustain when the pressure raised up to $2500$~Pa. Typical images of the discharge are  shown in Fig.~$2$ for different pressure levels. 
\begin{figure}[ht]
\begin{center}$
\begin{array}{ll}
(a)\\
\includegraphics[width=0.43\textwidth,angle=0]{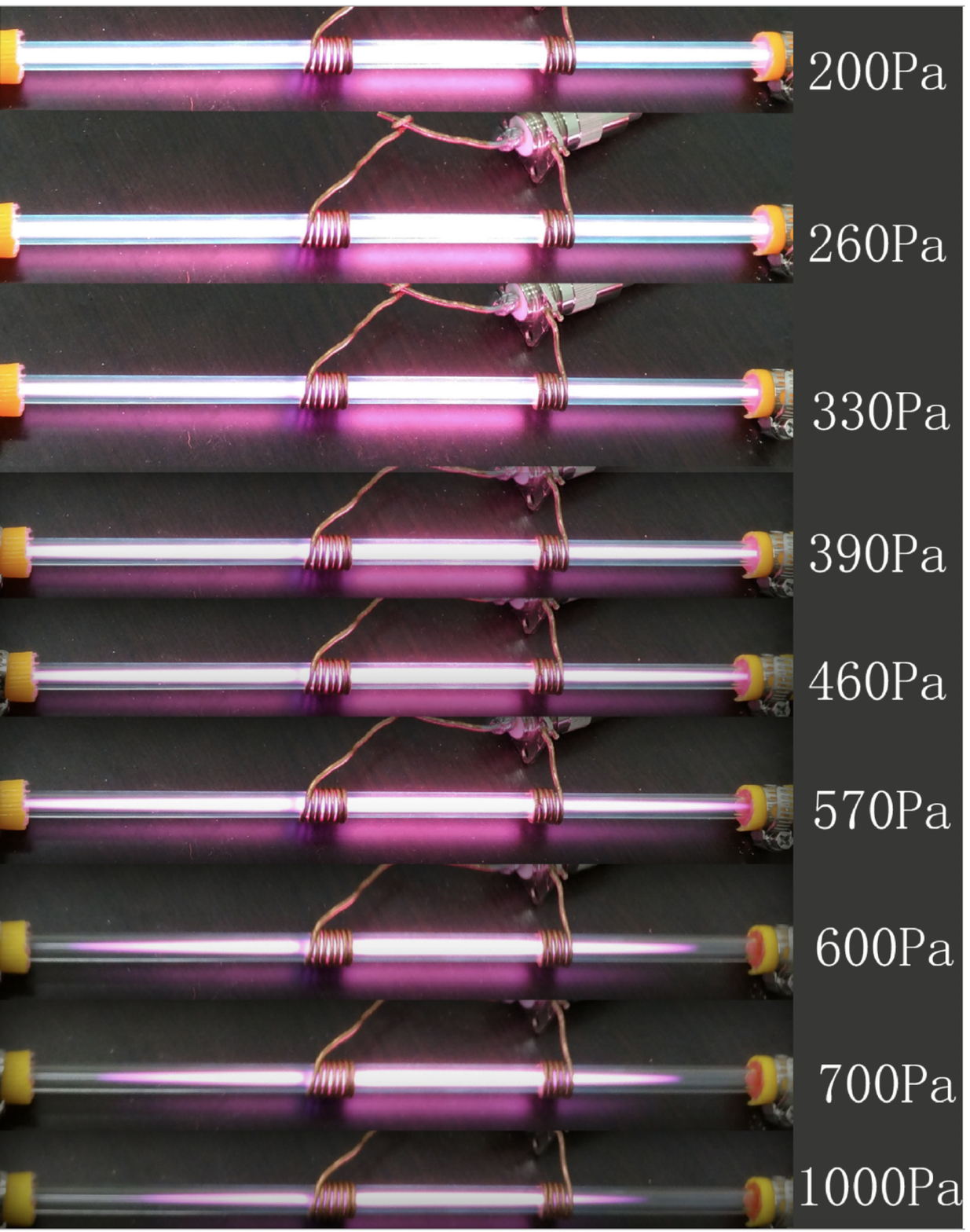}\\
(b)\\
\includegraphics[width=0.43\textwidth,angle=0]{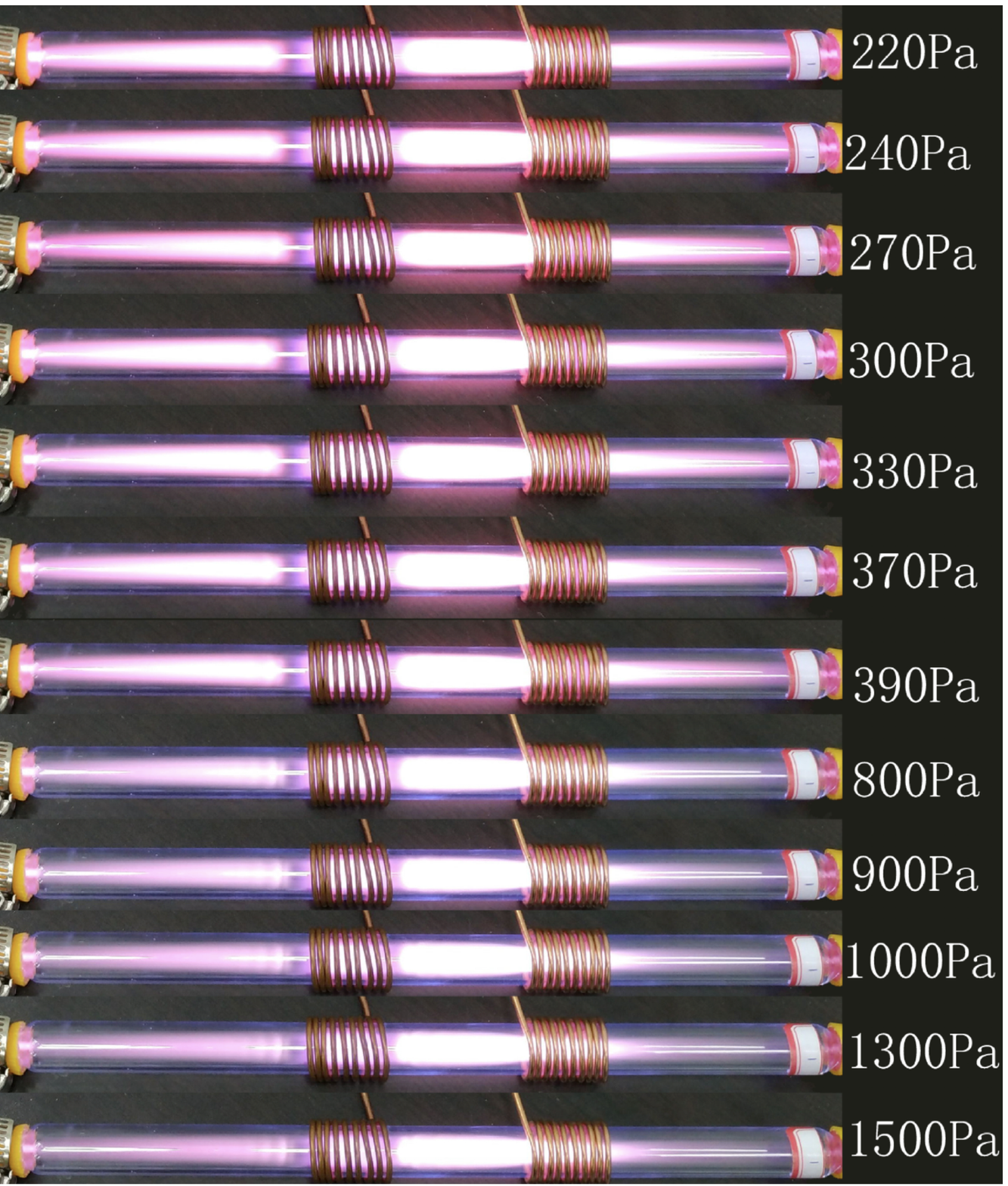}
\end{array}$
\end{center}
\caption{Typical images of discharge for various pressure levels: (a) inner diameter of quartz tube is $5$~mm, (b) inner diameter of quartz tube is $12$~mm.}
\label{fg2}
\end{figure}
We can see that the discharge is brighter for lower pressure, indicating higher ionization rate and plasma density there, and the area between two solenoids is always brighter than other areas, which could be attributed to axial confinement from magnetic mirrors and the resulted high ionization rate. Moreover, an asymmetric preference can be seen in the axial direction especially at high pressure levels, probably caused by the twist direction of solenoid relative to the voltage drop. In the fact, this twist direction, which determines the direction of induced magnetic field and thereby the dielectric property of entire plasma, is our main focus of ongoing research to optimize the discharge efficiency and maximize the asymmetric plume for propulsion purpose. Typical wave form of voltage drop across the isolated solenoids is shown in Fig.~$3$ for quartz tube in diameter of $5$~mm and pressure of $700$~Pa.
\begin{figure}[ht]
\begin{center}
\includegraphics[width=0.5\textwidth,angle=0]{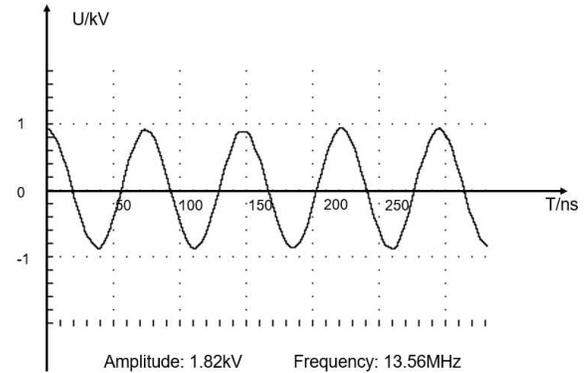}
\end{center}
\caption{Typical wave form of voltage drop across the isolated solenoids from oscilloscope measurement.}
\label{fg1}
\end{figure}

This efficient ionization method is particularly attractive for plasma propulsion in near space, and has three advantages over other techniques such as dielectric barrier discharge. First, there is no limitation on the loading voltage between electrodes, because it is free of dielectric breakdown; second, the antenna has no contact with working gas so that it is free of plasma sputtering and thereby has a long life; third, this is an air-breathing propulsion system and there is no narrowing-down throat or slowing-down grid for air flows, therefore, the flow rate and hence propulsion efficiency can be very high. Other applications may spread into fields such as plasma-material processing, plasma biomedicine, plasma flow control and plasma-assisted ignition and combustion, where pressure in range of tens to thousands of pascals is employed.

\section*{Acknowledgment}
This work is supported by various funding sources: National Natural Science Foundation of China (11405271), China Postdoctoral Science Foundation (2017M612901), Chongqing Science and Technology Commission (cstc2017jcyjAX0047), Chongqing Postdoctoral Special Foundation (Xm2017109), Fundamental Research Funds for Central Universities (YJ201796), Pre-research of Key Laboratory Fund for Equipment (61422070306), and Laboratory of Advanced Space Propulsion (LabASP-2017-10).

\end{document}